\documentstyle[prd,aps,epsfig,tighten,12pt]{revtex}
\begin{document}
%\preprint{                                            Ref.~SISSA-15-99-EP}
\draft
%--------------------------------------------------------------------------
\title{     Can unstable relics save pure Cold Dark Matter ?    }
\author{	A.\ Masiero, D.\ Montanino, and M.\ Peloso	}
\address{
	Scuola Internazionale Superiore di Studi Avanzati (S.I.S.S.A.),\\
		Via Beirut 4, I-34014 Trieste, Italy\\
			        and\\
	Istituto Nazionale di Fisica Nucleare, Sez.\ di Trieste\\
		Via Valerio 2, I-34127 Trieste, Italy
}\maketitle
%--------------------------------------------------------------------------
\vspace{0.8cm}
\begin{abstract}
The standard CDM model fails to describe the power spectrum of
fluctuations since it gives too much power at small scales. Among other
possible improvements, it has been suggested that an agreement with
observations can be achieved with the addition of a late decaying
particle, through the injection of non-thermal radiation and the
consequent increase of the horizon length at the equivalence time. We
analyze the possibility of implementing this idea in some extensions of
the electroweak standard model, discussing the cosmological and
astrophysical bounds to which these schemes are subject.
\end{abstract}
 % Particle theory of the early Universe
%--------------------------------------------------------------------------

\vspace{1.8cm}

\thispagestyle{empty}

%--------------------------------------------------------------------------
\section{Introduction}
%--------------------------------------------------------------------------

The topic of structure formation in the early universe is still an open
problem which has challenged many cosmologists in the past years. In the
last decade great attention was devoted to the standard Cold Dark Matter
(CDM) model, characterized by a flat universe with cold particles as dark
matter, and by density fluctuations evolving from a primordial scale
invariant (Harrison-Zel'dovich) spectrum.

This model turned out quite successful and its predictions about the
properties of the different kinds of galaxies that form agreed well with
observations \cite{bfpr}. However, more recent results \cite{defw,fdsyh}
showed that, once normalized at large scale by COBE data, the standard CDM
model predicts more power than observed at small scale. Several ideas have
been proposed to improve the situation. A possibility is to change the
inflationary prediction on the primordial spectrum.  A non flat spectrum
can be achieved for example in supergravity inflationary models
\cite{ars}, in models with more than one inflaton \cite{sbb}, or also by
considering a non adiabatic evolution of the inflaton field \cite{lps}. 
Another possibility is offered by mixed hot plus cold dark matter
\cite{mdm} and by ``warm'' dark matter models \cite{warm}, since light
particles have high speed at the decoupling era and cannot cluster on
small scales.

Another way to fit the observations is the introduction of a non-zero
cosmological constant $\Lambda$ \cite{esm}. At present this is the most
appealing solution, since an incoming agreement about a cosmological
constant is being provided by supernovae measurements \cite{scp,sst}. By
examining high-redshift SNe Ia and using them as candles to determine the
distances to faraway galaxies, it is indeed possible to find evidence for
an accelerating universe and thus exclude models with
$\Omega_{\,\Lambda}=0$. 

While statistics of these experiments is becoming more and more solid with
the increasing of the number of observed high-$z$ supernovae (42 for
\cite{scp} and 16 for \cite{sst}), still --- as the authors of
\cite{scp,sst} say themselves --- some doubts can be cast on their
systematics.~\footnote
%--------------------------------------------------------------------------
{It is worthwhile to remark here that the introduction of a cosmological
constant opens many other unsolved puzzles. First of all, if one
associates the cosmological constant to the energy of the vacuum, one
would expect either $\Omega_{\,\Lambda}=0$ (if some symmetry prevents it)
or $\Omega_{\,\Lambda}\sim M_{\,{\rm pl}}^4/\rho_{\, {\rm critical}}$,
while $\Omega_{\,\Lambda}={\rm O}(1)$ appears as an incredibly fine-tuned
small value. Second, it is not clear why $\Omega_{\,\Lambda}$ should be of
the same order of $\Omega_{\,{\rm matter}}$. In order to solve these
problems, quintessential models --- where $\Omega_{\,\Lambda}$ is a
dynamical parameter associated to the energy of a slow evolving scalar
field --- are becoming popular, but they are far from being conclusive
\cite{qui}.}
%--------------------------------------------------------------------------

Another point in favor of the option of a non-vanishing vacuum energy
contribution comes from measurements of $\:\Omega_{\rm matter}\:$ and 
$\:\Omega_{\rm total}\:=\:\Omega_{\rm matter}\:+\:\Omega_{\Lambda}$. The
latter quantity can be observationally determined from the position of the
first acoustic peak of the Cosmic Background Radiation (CBR). Preliminary
results (which are still plagued by large errors) indicate a preference
for $\:\Omega_{\rm total}\simeq 1$ (see \cite{tur} and references therein).
Clearly, also theoretical arguments based on inflation strongly favor a
flat universe. As for $\:\Omega_{\rm matter}\:$, its best determination
comes from measurements of the baryon fraction in clusters. Using clusters
as ``fair samples'' of matter in the universe, the bound on $\:\Omega_{\rm
baryons}\:$ from the standard nucleosynthesis, and determinations of the
amount of baryons and the total matter in clusters, it has been inferred
that $\:\Omega_{\rm matter}\:$ should not exceed 40\%
\cite{clu}.~\footnote
%--------------------------------------------------------------------------
{Other circumstantial evidences for $\:\Omega_{\rm matter}\:<\:1$ related
to the evolution of the abundance of rich cluster with redshift are more
controversial.}
%--------------------------------------------------------------------------
As convincing as this arguments look for $\:\Omega_{\rm matter}\:<\:1$, we
think that they should not be considered conclusive on the intriguing
issue of the matter present in the universe. 

In view of the present uncertainties, we think it is still valuable to
consider other options tackling the failure of the standard CDM model. In
this paper we focus on the possibility that, in addition to the CDM, there
exist decaying (or sometimes called volatile) particles. The decay
products of these particles can be used to increase the horizon length at
the equivalence time, hence increasing the power at large scales
\cite{bond,kim}.

In the past years, only a couple of schemes which implement this idea have
been proposed (see for example \cite{bond,kim,wgs,kima}). Indeed, the
candidate models must have among the decay products only very weakly
interacting particles, since stringent cosmological and astrophysical
bounds apply when more interacting particles are produced. The most
popular scenario for the realization of the shift of $\lambda_{\,{\rm
eq}}$ involves an $m \sim \,$MeV $\,$ neutrino decaying in majoron plus a
lighter neutrino. An alternative possibility is given by decays with the
presence of the axion field and its superpartners.

The aim of the present paper is to update and extend the above analyses in
light of some other candidate models and of the most recent observational
data. The main new possibility that we suggest involves decays with
production of high energy neutrinos. These processes must be handled with
care since they can enter in conflict with bounds imposed by the non
observations of the neutrinos (mainly for very late decays) or (for early
enough decays) by Big-Bang Nucleosynthesis. These last bounds apply
because the produced neutrinos can scatter off background antineutrinos,
generating an electromagnetic cascade which can photodissociate the light
elements previously produced.

The work is organized as follows. In Sec.\ II$\,$ we present the general
idea of the shift of the equivalence scale. In Sec.\ III$\,$ we report the
main cosmological and astrophysical bounds which apply in case of
production of high energy neutrinos. In Sec.\ IV$\,$ we discuss the
possibility to implement this idea in a supersymmetry context. In Sec\
V$\,$ we switch to the option involving $m \sim\,$ MeV $\,$ unstable
neutrinos and we study two other decaying processes (namely the decay
into three neutrinos and into neutrino plus familon) in addition to the
more common decay into neutrino plus majoron. The results and some
prospects for the future are presented in the conclusions.

%--------------------------------------------------------------------------
\section{Shift of the equivalence scale}
%--------------------------------------------------------------------------

As we said in the introduction, the standard CDM model fails to describe
the observed spectrum of fluctuations. However, an agreement with
observations can still be reached in an $\Omega_{\, {\rm CDM} \,+\, {\rm
baryons}} = 1$ context if the horizon length at the equivalence scale is
increased. The reason is that during the radiation dominated era super
horizon fluctuations can grow, while sub horizon ones can not. Thus, if we
delay the matter domination (or, in other words, we increase the horizon
length $\lambda_{\,{\rm eq}}\:$) we give the large modes more time for
growing, while the small (sub horizon)  ones are kept frozen by a sort of
``pressure of radiation''.

In standard models (with relativistic particle at present provided by CBR
photons and three species of relic neutrinos) the length of the horizon at
the equivalence time is given by
%--------------------------------------------------------------------------
\begin{equation} \label{leq}
\lambda_{\,{\rm eq}} \simeq 30 \left( \Omega \, h^2 \right)^{-\,1}
\:{\rm Mpc}
\;\;\;,\end{equation}
%--------------------------------------------------------------------------
where $\Omega$ is the ratio density of the universe over critical density
($\Omega = 1$ for a flat universe) and $h$ is the Hubble parameter in
units of $100 \;{\rm Km}\;{\rm sec}^{-\,1}\:{\rm Mpc}^{-\,1}\:$. 
Inflationary models strongly prefer $\Omega = 1\;$, while observations 
constrain $h \:\in \left[ 0.4 \:,\: 1 \right]\:$.~\footnote
%--------------------------------------------------------------------------
{More recent data suggest to tighten this interval for $h$ ($\,h = 0.67 \pm
0.10\,$ \cite{hubble}). However, since a stable agreement on the value
of the Hubble constant has not been reached so far, in our analysis we
prefer to stick to the more conservative range $\left[ 0.4 \:,\: 1 
\right]\,$.}
%--------------------------------------------------------------------------
On the other hand, the observed spectrum of fluctuations is well fitted by 
CDM models only for
%--------------------------------------------------------------------------
\begin{equation} \label{leqvol}
\lambda_{\,{\rm eq}} \simeq 150 \: h^{-\,1} \:{\rm Mpc}
\end{equation}
%--------------------------------------------------------------------------
i.e., for $\Omega \, h = 0.2\;$.

Instead of lowering $\Omega \, h\;$, a raising of $\lambda_{\,{\rm eq}}$
can be achieved by considering an unstable matter which decays into
relativistic particles \cite{bond,kim}. Photons cannot accomplish this
task:  indeed $\gamma$ produced after $t \simeq 10^6 \:{\rm sec}$ must be
of negligible amount or they would distort the shape of the CBR (see next
section), while photons emitted at early times thermalize and just
contribute to rising the temperature of the (thermal) background to the
value we measure today [and from which we get Eq.~(\ref{leq})]. If the
decay products are not photons (let us call them ``invisible''), the
result is that the relativistic energy is greater than the one that we
infer just measuring the temperature of the CBR. Thus, for appropriate
choice of the decay time $\tau\;$, $\lambda_{\,{\rm eq}}\,$ can increase. 

%==========================================================================
\begin{figure}[h!]
\begin{center}
\epsfig{bbllx=50pt,bblly=280pt,bburx=580pt,bbury=750pt,height=14truecm,
        figure=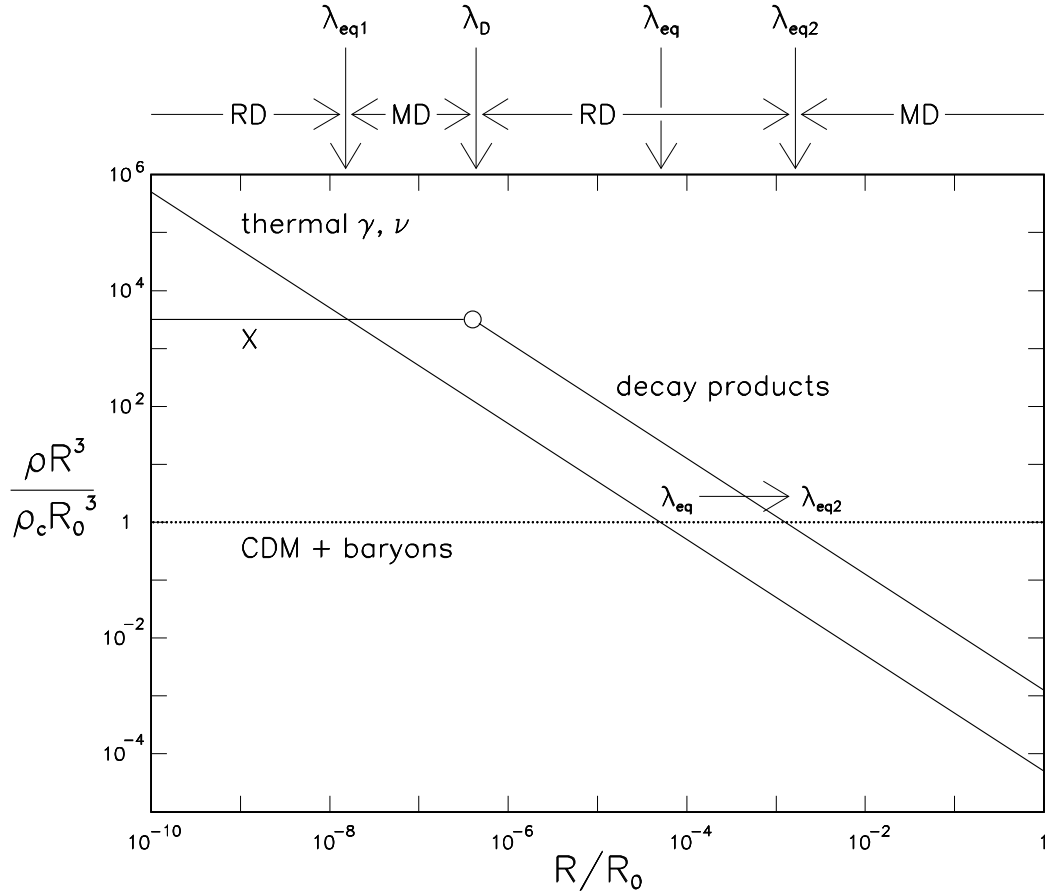}
\caption{\label{fig1}
\footnotesize Schematic representation of the model. RD and MD denotes,
respectively, Radiation and Matter Domination. See the text for other
details.}
\end{center}
\end{figure}
%==========================================================================

In Fig.~\ref{fig1} (inspired by the work of H.B.\ Kim and J.E.\ Kim 
\cite{kim}) we give a schematic representation of the model that we have
described.

We denote by $X$ the matter which decays, by $m\:Y_{\,{\rm tot}}$ its 
(comoving) energy density,~\footnote
%--------------------------------------------------------------------------
{Following the standard notation, $m$ is the mass of the particle, while
$Y_{\,{\rm tot}}$ is the ratio between its number density and the entropy
density of the universe.}
%--------------------------------------------------------------------------
and by $m\:Y_{\,{\rm rel}}\,$ the part of it which
goes into relativistic particles.~\footnote
%--------------------------------------------------------------------------
{For example, in Sec.\ III$\,$ we will discuss the decay gravitino
$\rightarrow$ sneutrino $+$ neutrino. The sneutrinos produced by this
process have a mass which is typically of the same order of the one of the
gravitinos, and thus they quickly derelativizes. Hence, the only energy
which contributes to the shift of $\lambda_{\,{\rm eq}}$ and which we
denote by the suffix $\,rel\,$ is the one associated to the neutrinos.
Thus, in this case $m\:Y_{\,{\rm rel}}\, \sim \, m\:Y_{\,{\rm tot}}\,$.}
%--------------------------------------------------------------------------

The $X$ particle is not relativistic before decaying, thus its energy
density increases with respect to the thermal one. In order to realize the
shift of $\lambda_{\,{\rm eq}}\:$, we see from Fig.~\ref{fig1} that $X$
must decay only after it dominates. Thus the main peculiarities of this
scheme are (i) that $\lambda_{\,{\rm eq}} \rightarrow \lambda_{\,{\rm
eq2}} > \lambda_{\,{\rm eq}}\:$, and (ii) that a new matter domination era
--- starting at $\lambda_{\,{\rm eq1}}\:$ --- arises. The values of these
two  lengths are \cite{kim}
%--------------------------------------------------------------------------
\begin{eqnarray}
\lambda_{\,{\rm eq1}} &\simeq& 8 \cdot 10^{-2} \:{\rm Kpc}\: \left(
\frac{{\rm MeV}}{m\:Y_{\,{\rm tot}}} \right) \nonumber\\
\lambda_{\,{\rm eq2}} &\simeq& 30 \:{\rm Mpc}\: \left( \Omega \, h^2
\right)^{-\,1} \: \left[ \left\{ \frac{1}{0.55} \left( \frac{\tau}{
{\rm sec}}\right) \: \left( \frac{m\:Y_{\,{\rm rel}}}{{\rm MeV}} 
\right)^2\right\}^{2/3} + 1 \right]^{1/2} \nonumber
\;\;\;.\end{eqnarray}
%--------------------------------------------------------------------------

The scheme is implemented if $\lambda_{\,{\rm eq2}} \simeq 150 \: h^{-\,1}
\:{\rm Mpc}$ [cfr. Eq.~(\ref{leqvol})], namely:
%--------------------------------------------------------------------------
\begin{equation} \label{kimeq}
\left( \frac{\tau}{{\rm sec}} \right) \: \left( 
\frac{m\:Y_{{\rm rel}}}{{\rm MeV}} \right)^2 \simeq 0.55 \: \left[
\left( h/0.2 \right)^2 - 1 \right]^{3/2}
\;\;\;.\end{equation}
%--------------------------------------------------------------------------

For the model to phenomenologically hold, some bounds have to be
respected. Those stated here are somehow model independent and represent
general constraints that have always to be taken into account.  In the
next section we will instead discuss limits which apply in case of
production of high energy neutrinos. 

The first bound comes from nucleosynthesis and requires that at the time
of neutrino decoupling the energy density of $X$ is less than that of one
neutrino species \cite{lisi}:
%--------------------------------------------------------------------------
\begin{equation} \label{upplim}
\left( \frac{m\:Y_{\,{\rm tot}}}{{\rm MeV}} \right) < 0.107
\;\;\;.\end{equation}
%--------------------------------------------------------------------------
The demand that the fraction of the energy density into relativistic decay
products is high enough to realize the shift yields: 
%--------------------------------------------------------------------------
\begin{equation} \label{lowlim}
\left( \frac{m\:Y_{\,{\rm rel}}}{{\rm MeV}} \right) > 
3.6 \cdot 10^{-\,6} \: h^2
\;\;\;.\end{equation}
%--------------------------------------------------------------------------
Notice that imposing Eqs.~(\ref{kimeq}), (\ref{upplim}), and
(\ref{lowlim}), we safely avoid overclosure of the universe by the decay
products.

%--------------------------------------------------------------------------
\section{Cosmological and astrophysical bounds}
%--------------------------------------------------------------------------

As already remarked, the bounds considered in the previous section are
certainly necessary, but they can be not sufficient if the decay products
affect our present observations in some other ways than the only shift of
the equivalence scale.~\footnote
%--------------------------------------------------------------------------
{The bounds~(\ref{upplim}) and~(\ref{lowlim}) are sufficient for processes
like axino $\rightarrow$ axion $+$ gravitino \cite{kim}, or for decays
with production of low energy neutrinos, since all these particles have
very weak interactions.}
%--------------------------------------------------------------------------

For example, severe limits exclude the possibility of realizing the shift
of $\lambda_{\rm eq}$ if the decay produces a non negligible amount of
photons.~\footnote
%--------------------------------------------------------------------------
{The authors are grateful to S.\ Bonometto for remarking this point.}
%--------------------------------------------------------------------------
Indeed, Eq.~(\ref{kimeq}) is satisfied only if the relativistic energy
density is about $25$ times \cite{kim} larger than the thermal one (that
is the one that we infer by measuring the temperature of the microwave
background radiation). As a consequence, if the decay occurs at early
time, the produced photons thermalize and thus must not carry more than
$1/25\,$ of the total energy density of the decaying particles. On the
other hand, if the decay happens later the photons produced must be much
less than the thermal ones, or they would appear as a clear distortion in
the observed black-body spectrum. In both cases, it follows that the
amount of energy density in form of photons must be much less than the one
undergone into other relative particles.

From the above discussion we realize that the shift of $\lambda_{\rm eq}$
cannot be implemented in decays with the production of photons. We thus
switch to the case of high energy neutrinos among the decay products,
which needs a more detailed discussion. In fact, in this case one has to
discuss some limits which come from the non observation of the neutrinos
and from BBN.

\begin{itemize}

\item
Non observation: The limits coming from the IMB nucleon decay detector are
low enough \cite{eglns,ggs}~\footnote
%--------------------------------------------------------------------------
{See also \cite{ber}, where the specific case of a decaying fermion with
abundance determined by annihilation via Standard Model gauge bosons is
considered.}
%--------------------------------------------------------------------------
to rule out a flux of neutrinos necessary to shift $\lambda_{\,{\rm
eq}}$ if they have at present an energy $E_{\nu_0} > 100\:$MeV.

The energy interval $1.5 \:{\rm MeV}\: < E_{\nu_0} < 100 \:{\rm MeV}\:$
has been in the past or it is currently monitored by several experiments 
\cite{kam2,lvd,supkam} and no claim has been made about an anomalous
diffuse neutrino flux.

We thus impose the safe requirement $E_{\nu_0} < 1 \, {\rm MeV} \,$, i.e.
that neutrinos coming from the decay of the X matter do have a present
energy lower than $1 \,{\rm MeV}\,$.

Making the approximation $ \; E_{\nu_0} \simeq \left( \tau /
t_{\,{\rm eq}} \right)^{1/2}\:\left( t_{\,{\rm eq}} / t_0 \right)^{2/3}
\: E_\nu \;,$ where $E_\nu$ is the energy of the neutrinos at their
production, we get
%--------------------------------------------------------------------------
\begin{equation} \label{neuobs}
\frac{E_\nu}{{\rm GeV}} < 4 \cdot 10^6 \: \left( \frac{{\rm sec}}{\tau}
\right)^{1/2}
\;\;\;.\end{equation}
%--------------------------------------------------------------------------

We see that, for a given value of energy density of the produced neutrinos
$m\:Y_\nu\;$, decays with lower values of the energy of each neutrino (and
higher numerical density) are less constrained.

\item
BBN : neutrinos produced at early times can annihilate off
background antineutrinos and produce $e^\pm$ pairs; then the
charged particles can photodissociate --- via electromagnetic cascade ---
the light elements produced during BBN \cite{km,gss}.

In the limit of Fermi approximation~\footnote
%--------------------------------------------------------------------------
{Valid for $E_\nu \: E_{\bar \nu} < M_Z^2 \;$, i.e., $$ \left(
\frac{E_\nu}{{\rm GeV}} \right) \: \left( \frac{{\rm sec}}{\tau}
\right)^{1/2} < 10^7 \;\;\;,$$ that is for all cases of our interest.}
%--------------------------------------------------------------------------
the fraction of neutrinos ``converted'' into radiation is \cite{gss}
%--------------------------------------------------------------------------
\begin{equation} \label{fneurad} f_{\,\nu \rightarrow \gamma} \simeq 1.3
\cdot 10^{-\,3} \: \left( \frac{E_\nu}{{\rm MeV}} \right) \: \left(
\frac{{\rm sec}}{\tau} \right)  \;\;\;.\end{equation}
%--------------------------------------------------------------------------
Thus, once the neutrinos from the decay have enough energy to generate
$e^\pm$ pairs,~\footnote
%--------------------------------------------------------------------------
{i.e., for $$\left( \frac{E_\nu}{{\rm MeV}} \right) \: \left(
\frac{{\rm sec}}{\tau} \right)^{1/2} > 0.3 $$}
%--------------------------------------------------------------------------
we must calculate the density of radiation produced 
%--------------------------------------------------------------------------
\begin{equation}
\left( m\:Y \right)_\gamma = \left( m\:Y \right)_\nu \,\cdot\, 
f_{\,\nu\rightarrow\gamma}\;\;,
\end{equation}
%--------------------------------------------------------------------------
and analyze if it can spoil the good agreement between the observational
data and the Standard Big-Bang Nucleosynthesis (that is in absence of
decaying particles). In this way we can get an upper limit on $f_{\,\nu
\rightarrow \gamma}$ as a function of the decay lifetime $\tau\,$.

We do not review here the whole analysis about the impact of an
electromagnetic cascade on the abundance of the light elements, but we
just state the results of the numerical work \cite{hkkm}, addressing the
interested reader to the references listed there.~\footnote
%--------------------------------------------------------------------------
{See also \cite{psb}, where a detailed analysis including electromagnetic
cascade has been performed for lifetime bigger then $10^{15}$ sec.}
%--------------------------------------------------------------------------

The two primordial abundances that most constrain the theoretical
predictions are the $^4$He and the deuterium ones. At present , different
observations give the amount of these abundances with some discrepancies:
%--------------------------------------------------------------------------
\begin{eqnarray}
\text{Low D}\;\;:\;\; y_2 &=& \left( 3.39 \pm 0.25 \right) \cdot
10^{-\,5} \;\;,\;\; y_2 = n_D/n_H \nonumber\\
\text{\rm High D}\;\;:\;\; y_2 &=& \left( 1.9 \pm 0.5 \right) \cdot
10^{-\,4} \nonumber\\
\text{Low $^4$He}\;\;:\;\; Y &=& 0.234 \pm \left( 0.002 
\right)_{{\rm stat}} \pm \left( 0.005 \right)_{{\rm syst}} \;\;,\;\; Y
= \frac{\rho_{\:^4{\rm He}}}{\rho_{\:{\rm baryons}}} \nonumber\\
\text{High $^4$He}\;\;:\;\; Y &=& 0.244 \pm \left(0.002
\right)_{{\rm stat}} \pm \left( 0.005 \right)_{{\rm syst}} \nonumber
\end{eqnarray}
%--------------------------------------------------------------------------
We discuss the implications considering the different above ranges.

%==========================================================================
\begin{figure}[t!]
\begin{center}
\epsfig{bbllx=70pt,bblly=60pt,bburx=540pt,bbury=740pt,height=18truecm,
        figure=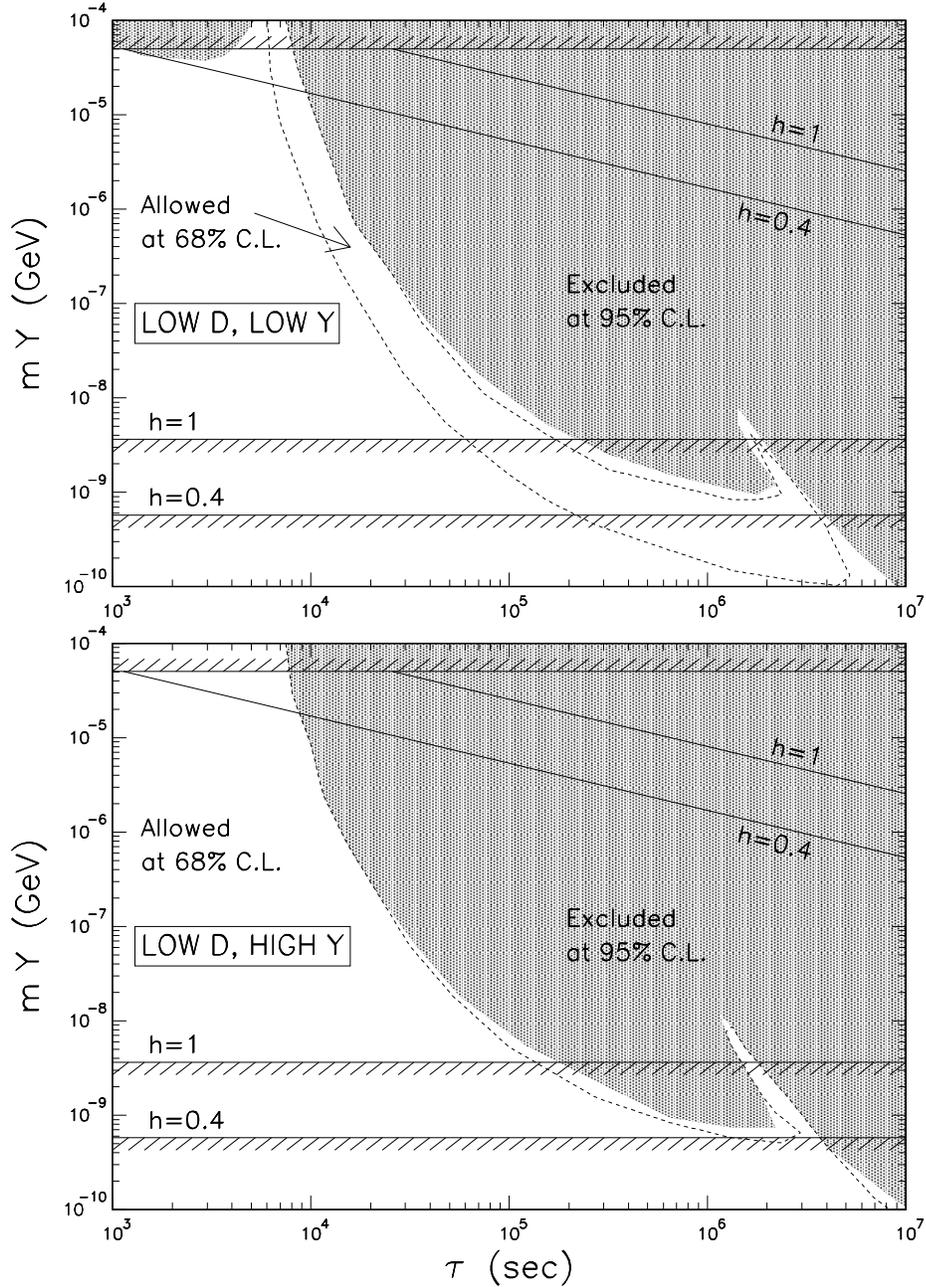}
\caption{\label{fig2}
\footnotesize Bounds on an electromagnetic cascade in the case of low D
and $^4$He (upper panel) low D and high $^4$He (lower panel). Gray area:
excluded zone at 95\% C.L. by nucleosynthesis; dotted line: limit at 68\%
C.L. by nucleosynthesis. See the text for details.}
\end{center}
\end{figure}
%==========================================================================

In the case of low D and low $^4$He, the predictions of the standard BBN
(i.e.\ without decaying particles) are excluded at $68\,\%$ confidence
level. A non-vanishing $m\:Y_\gamma$ is preferred in order to destroy a
sufficient amount of the primordial deuterium. Thus, if this case will
turn out to be the right one, it could be worth to examine if the scheme
that we are proposing here can be advocated both to realize the shift of
$\lambda_{\,{\rm eq}}$ and to explain the would be ``standard BBN
failure''. 

The other three cases give qualitatively similar results. For all of them
standard BBN is allowed and we have upper bounds on $m\:Y_\gamma$ for each
value of $\tau\,$.

The results are summarized in Fig.~\ref{fig2}. First we report (from
\cite{hkkm}) the bounds imposed by BBN on $m\:Y_\gamma\,$. We show only
the case of low D and low $^4$He and the case of low D and high $^4$He
(the other two being similar to the last one). The gray area denotes the
points in the plane $\:m\:Y_\gamma \;{\rm vs}\; \tau\:$ excluded at
$95\,\%$ C.L. The region within the dotted lines corresponds to points
allowed at $68\,\%$ C.L. (in the lower panel these regions extends over
the entire blank space). 

Then we analyze the impact on these regions of the implementation of the
shift of $\lambda_{\,{\rm eq}}$ with a decay in high energy neutrinos with
the subsequent production photons. The straight lines refer to a decay in
which the initial energy density is equally shared between neutrinos and a
massive decay particle which quickly derelativises ($\:m\:Y_\nu =
m\:Y_{{\rm rel}} = \frac{1}{2}\:m\:Y_{{\rm tot}}\;$). The horizontal lines
correspond to the upper and lower bounds $m\:Y_{{\rm tot}} < 0.107\:  {\rm
MeV}\;$ [Eq.~(\ref{upplim})] and $m\:Y_{{\rm \nu}} > 3.6 \cdot 10^{-\,6}
h^2\:{\rm MeV}\;$ [Eq.~(\ref{lowlim})], while the two diagonal ones
indicate for each $\tau$ the amount of $m\:Y_\nu$ produced when the shift
of $\lambda_{\,{\rm eq}}$ is realized [Eq.~(\ref{kimeq})]. 

The results shown in Fig.~\ref{fig2} put a severe constraint on $f_{\,\nu
\rightarrow \gamma}\,$, since they show that the amount of neutrinos
``converted'' in photons must be a very small fraction of the total one in
order to respect the limits imposed by BBN. For example, we see that at
$\tau = 10^7 \:\mbox{sec}\,$ the quantity $f_{\,\nu \rightarrow \gamma}$
must be smaller than about $10^{-\,5} \div 10^{-\,4}\:$ (according to
different values of $h\,$).~\footnote
%--------------------------------------------------------------------------
{The same limit must be imposed also on the branching ratio of the decay
channels in which photons are produced.}
%--------------------------------------------------------------------------

In the next section we study if it is possible to implement the shift of
$\lambda_{\rm eq}$ in supersymmetric models with and without $R$ parity.
The main part is devoted to the study of two processes with the production
of high energy neutrinos. We show there that the only one of the two model
which is workable satisfies also the bounds discussed in this section.

\end{itemize}

%--------------------------------------------------------------------------
\section{Supersymmetric decays}
%--------------------------------------------------------------------------
In supersymmetric models, the terms
%--------------------------------------------------------------------------
\begin{equation} \label{Rvio}
\lambda_{ijk}\,L_i\,L_j\,E_k^c \:+\: \lambda_{ijk}^{'}\,L_i\,Q_j\,D_k^c
\:+\: \mu_i\,L_i\,H_u \:+\: \lambda_{ijk}^{''}\,U_i\,D_j\,D_k^c
\end{equation}
%--------------------------------------------------------------------------
are usually dropped away from the superpotential in order to avoid fast
proton decay. To do this, a discrete symmetry which is called $R$ parity
and  which is violated by the above terms is imposed.

$R$ parity forbids terms in which supersymmetric particles are in odd
number. As a consequence, superparticles can be produced or destroyed only
in pairs, and the Lightest SuperParticle (LSP) is stable. An advantage of
supersymmetry is that this particle, for adequate but natural choices of
the Minimal Supersymmetric Standard Model (MSSM) parameters, is a good cold
dark matter candidate~\cite{jkg}.

However, proton decay requires the violation of both baryon and lepton
number. Thus it is not in disagreement with the present limits on proton
lifetime to make vanishing the last term in Eq.~(\ref{Rvio})  while
keeping different from zero some (or all) of the first three, or viceversa
to take only the last term non-vanishing. 

Another phenomenologically acceptable possibility is to break $R$ parity
spontaneously. In these scenarios one typically introduces some
isosinglets which acquire a $VEV$ and break $R$ and Lepton symmetries,
without entering in conflict with the bounds imposed by the $Z^0$ width.

Whatever possibility one may choose, as a result in both cases the LSP is
non longer stable and, unless we take very a very small breaking of $R$
\cite{bjv}, it can not be invoked as CDM. Nevertheless, it may be possible
for its decay to implement the shift of $\lambda_{\,{\rm eq}}\,$.

In the next two subsections we discuss the possibility of realizing the
shift of $\lambda_{\rm eq}$ in a supersymmetric context, first in schemes
with conserved $R$ parity and then in others where this symmetry is
broken.

%--------------------------------------------------------------------------
\subsection{Conserved $R$ parity}
%--------------------------------------------------------------------------

In the context of supersymmetry with conserved $R$ parity one would like
to implement the shift of the equivalence scale through the decay of the
next LSP to the LSP. The main difficulty is that ``ordinary''
superparticles decay extremely fastly. For example, the lifetime of the
decay ${\tilde \nu} \rightarrow {\tilde B} \; \nu$ is
%--------------------------------------------------------------------------
\begin{equation}
\tau \sim 10^{-\,24} {\rm  sec } \left( 1 - \frac{m_{\tilde B}^2}
{m_{\tilde \nu}^2} \right)^{-\,2}
\;\;\;,\end{equation}
%--------------------------------------------------------------------------
and a strong fine tuning is required in order to have a late decay.

It is  thus mandatory to consider processes with very low interacting
particles. One of these particles is the gravitino (that is the spin $3/2$
superpartner of the graviton) since its interactions are suppressed by
inverse powers of the Planck mass.

This option is already present in \cite{kim}, where the decay axino
$\rightarrow$ axion $+$ gravitino is considered. This scheme implements
the shift of $\lambda_{\,{\rm eq}}\,$ for several values of the axino
decoupling temperature with respect to the reheating temperature, and its
good ``flexibility'' is essentially due to the very little danger that
both the produced particles have on our observations. 

Another interesting scheme involving an unstable gravitino is present in
\cite{fg}, where the scenario characterized by the process~\footnote
%--------------------------------------------------------------------------
{We do not consider here other decay modes of the gravitino into
``ordinary'' superparticles, since they typically produce photons.}
%--------------------------------------------------------------------------
gravitino $\rightarrow$ neutrino $+$ sneutrino is discussed in details,
but without considering the problem of the shift of the equivalence scale.

The scheme proposed in \cite{fg} is a very interesting one, since it has
the following advantages:

\begin{itemize}

\item
In standard models either $\,m_{3/2} \leq \:$ keV $\:$ (if stable) or
$\,m_{3/2} \geq 10 \:$ TeV $\:$ (if it decays, this bound is imposed by
BBN).  Here the primordial gravitinos are diluted away by inflation, and
the ones considered are generated by the reheating. For appropriate values
of the reheating temperature, intermediate masses are allowed \cite{bbp}. 

\item
Usually the sneutrino is not assumed as the LSP, since its relic abundance
is very small. In this case we have instead $Y_{\tilde \nu} = Y_{3/2} \:$.

\item 
There is absence of direct $\gamma$ production.

\end{itemize}

The following relations hold \cite{bbp}:

\begin{itemize}

\item abundance from the reheating:
%--------------------------------------------------------------------------
\begin{equation} \label{gra1}
m\:Y_{3/2} = 2.6 \cdot 10^{-\,8} \:{\rm GeV}\: \left( \frac{T_r}{10^{13}
\:{\rm GeV}\:} \right) \: \left( \frac{m_{3/2}}{100 \:{\rm GeV}\:}\right) 
\;\;\; ;\end{equation}
%--------------------------------------------------------------------------

\item decay lifetime:
%--------------------------------------------------------------------------
\begin{equation} \label{gra2}
\tau = 3.92 \cdot 10^8 \:{\rm sec}\: \left( \frac{100 \:{\rm GeV}\:}
{m_{3/2}} \right)^3 \: \left( 1 - \frac{m_{\tilde \nu}^2} {m_{3/2}^2}
\right)^{-\,4} 
\;\;\; ;\end{equation}
%--------------------------------------------------------------------------

\item Fraction of energy transferred to the neutrino:
%--------------------------------------------------------------------------
\begin{equation} \label{gra3}
F = \frac{1}{2} \: \left( 1 - \frac{m_{\tilde \nu}^2} {m_{3/2}^2} \right)
\;\;\; ;\end{equation}
%--------------------------------------------------------------------------

\item Closure by $\: \tilde{ \nu}\: $:
%--------------------------------------------------------------------------
\begin{equation} \label{gra4}
\left( \frac{m_{\tilde \nu}}{100 \:{\rm GeV}\:} \right) \: \left( 
\frac{T_r}{10^{13} \: {\rm GeV}\:} \right) = 0.14 \: h^2 \:
\Omega_{\tilde \nu}
\;\;\; .\end{equation}
%--------------------------------------------------------------------------

\end{itemize}
In these equations, $T_r$ is the reheating temperature and $\Omega_{\tilde
\nu}$ is the contribution to the critical density of the sneutrinos from
the decay. We ask for $\Omega_{\tilde \nu} \simeq 1\;$.

The conservative experimental bound $m_{3/2} > m_{\tilde \nu} \geq 43 \:
{\rm GeV}$ \cite{efos} limits [from Eq.~(\ref{gra4})] $T_r$ and thus the
abundance $m\:Y_{3/2}\;$. It follows that, when we try to implement
Eq.~(\ref{kimeq}), we must delay $\tau$ by a fine tuning $m_{\tilde \nu}
\simeq m_{3/2}\;$. However, in this case the fraction $F$ of energy to the
neutrinos become very low and the bound $m \: Y_{3/2} \:  F > 3.6 \cdot
10^{-\,6} \: h^2 \:{\rm MeV}\:$ is never satisfied (in other words, we
have neutrino domination only if we delay the decay after the allowed
time). 

We can ask how we can improve the situation. One possible way is to
require that the sneutrino is not the LSP and thus the ones produced by
gravitino decay go in turn into neutrino plus neutralino (now the LSP). In
this case we can expect an improvement since, even with fine tuning, the
second neutrino has about half the energy of the initial gravitino. 

In order for this scheme to work, the direct decay gravitino $\rightarrow$
neutralino $+$ its ordinary partner must be forbidden by kinematics, or it
would be the dominant channel (with consequent production of photons). The
best that we can do is to choose $\chi \: = \:$ pure higgsino and forbid
the direct decay by suitably rising the mass of the corresponding higgs 
particle.~\footnote
%--------------------------------------------------------------------------
{This mass must also be kept high in order to avoid the decay $G
\rightarrow {\tilde H} \, f \, {\bar f} \:$ via virtual higgs.}
%--------------------------------------------------------------------------
However, it turns out that this scheme requires $m_{3/2} = \:{\rm O}
\left( \,{\rm TeV}\, \right)$ and thus raising the higgs mass at this
scale appears rather unnatural.

A possible solution is to consider the inverse process with the sneutrino
as NLSP which decays into the gravitino plus neutrino. The first problem
is the possible overclosure of the universe by the relic gravitinos. Also
in this case we can overtake it by requiring that inflation dilutes away
the gravitinos from the big-bang and by appropriately bounding the
reheating temperature \cite{bbp}.

The second difficulty of this scenario comes from the low relic sneutrino
abundance, due to the high $Z^0 mediated$ annihilation cross section.
However, this relic density increases in presence of an asymmetry between
the sneutrinos and the antisneutrinos (this is possible since sneutrinos
and antisneutrinos can be distinguished by their opposite lepton number).
Indeed in this case almost only the sneutrinos (or the antisneutrinos) in
excess survive the annihilation.~\footnote
%--------------------------------------------------------------------------
{The other annihilation channel ${\tilde \nu} \: {\tilde \nu} \rightarrow
\nu \: \nu$ through zino exchange must also be suppressed by suitably
rising the mass of the mediator.}
%--------------------------------------------------------------------------
In this way one can consider the sneutrino relic abundance $Y_{\tilde
\nu}$ as a free parameter dependent on the choice of the initial 
asymmetry.

The lifetime of this decay is given by \cite{live}
%--------------------------------------------------------------------------
\begin{equation}
\tau = 1.8 \cdot 10^8 \,\mbox{sec}\: \left( \frac{m_{3/2}}{10 \,\mbox{GeV}} 
\right)^2 \: \left( \frac{50 \,\mbox{GeV}}{m_{\tilde \nu}} \right)^5 \;.
\end{equation}
%--------------------------------------------------------------------------

Consequently, the shift of $\lambda_{\rm eq}$ is achieved for~\footnote
%--------------------------------------------------------------------------
{When we use this notation, the underlined value is referred to the $h=1$
case, while the other one to $h=0.4\;$.}
%--------------------------------------------------------------------------
\begin{equation}
Y_{\tilde \nu} = \left( 5.0 \cdot 10^{-\,9} \div 
\underline{2.4 \cdot 10^{-\,8}} \right) \: 
\left( \frac{10 \,\mbox{GeV}}{m_{3/2}} \right) \: 
\left( \frac{m_{\tilde \nu}}{50 \,\mbox{GeV}} \right)^{3/2} \;.
\end{equation}
%--------------------------------------------------------------------------

However, this value is too high, since the gravitinos produced by the
decay would contribute to the critical density by
%--------------------------------------------------------------------------
\begin{equation} \label{gravclos}
\Omega_{3/2} = \frac{m_{3/2} \, Y_{\tilde \nu} \, s^0}{\rho_{\rm cr}^0}
\simeq \left( 90 \div \underline{70} \right) \: \left( \frac{m_{\tilde
\nu}}{50 \, \mbox{GeV}} \right)^{3/2} \;. 
\end{equation}
%--------------------------------------------------------------------------
In Eq.~(\ref{gravclos}) the dependence of $\Omega_{3/2}$ on the gravitino
mass cancels out and one is left with the only free parameter $m_{\tilde
\nu}\,$. Since $m_{\tilde \nu} \geq 43 \: {\rm GeV}$ \cite{efos} from
accelerator experiments, the scheme here described cannot work because
the produced gravitinos would overclose the universe. 

This discussion shows that the realization of the shift of $\lambda_{\rm
eq}$ in an $R$ conserving supersymmetric scenario is problematic, although
one has {\it a priori} the possibility to consider particles very low
interacting and thus with long enough lifetimes. In particular, finding
alternative schemes to the one involving gravitinos and axinos together
(as discussed in \cite{kim})  appears a very difficult task.

%--------------------------------------------------------------------------
\subsection{Broken $R$ parity}
%--------------------------------------------------------------------------

In models with broken $R$ parity one may try to implement the shift of
$\lambda_{\rm eq}$ through the decay of the LSP. Indeed, one can consider
very long lifetimes for this particle simply lowering enough the breaking
of the $R$ symmetry.

If the breaking is made via one (or more) of the soft terms listed in
Eq.~(\ref{Rvio}), the decay typically leads to a nonnegligible production
of photons or charged particles. As an example, if neutralino is the LSP
and if one breaks $R$ through the term $\lambda_{ijk}\,L_i\,L_j\,E_k^c$,
the process $\chi \rightarrow \nu\,e^+\,e^-$ (through sneutrino or
selectron exchange) constitutes the main decay channel.~\footnote
%--------------------------------------------------------------------------
{If $R$ is broken by another of the terms in Eq.~(\ref{Rvio}), also quarks
are produced in the decay.}
%--------------------------------------------------------------------------
The same problem also arises in Gauge Mediated Supersymmetry Breaking
(GMSB) models \cite{fay}, where the LSP is typically an $m \sim \,$keV
gravitino. In presence of soft breaking of $R$ parity the gravitino is
very long living and it typically decays via loop into photons and
neutrinos. We thus conclude that these schemes offer very little chances
to implement the shift of $\lambda_{\rm eq}\,$, since --- as we said above
--- decays with the production of photons must be excluded.

A completely different and more workable scenario is offered by the
possibility of breaking $R$ parity spontaneously. In the model described
in \cite{masval} this is achieved by the extension of the MSSM with three
isosinglet scalars~\footnote
%--------------------------------------------------------------------------
{The presence of all these new fields is necessary if one wants to 
break the lepton symmetry at tree level. However, if one allows for
breaking through radiative corrections, the number of these singlets can
be reduced \cite{gmpr}.}
%--------------------------------------------------------------------------
$\;\Phi\,$, $\nu^c\,$, and $S$ carrying, respectively, lepton number (L)
$\;0$, $-\,1$, and $1\,$.

In this model, the breaking of both global $L$ and $R$ symmetries gives
birth to a massless goldstone boson $J$ --- called majoron --- given by
the imaginary part of
%--------------------------------------------------------------------------
\begin{equation} \label{majsup}
\frac{v_l^2}{V\,v^2} \: \left( v_u \, H_u - v_d \, H_d \right) +
\frac{v_l}{V} \, {\tilde \nu} - \frac{v_r}{V} \, {\tilde \nu^c} +
\frac{v_s}{V} \, {\tilde S} \;\;.
\end{equation}
%--------------------------------------------------------------------------
In the above equation the $H$ fields are the two higgs fields, ${\tilde
\nu}$ is the sneutrino, $V = \sqrt{v_r^2 + v_s^2}\,$, $v = \sqrt{v_u^2 +
v_d^2}\:$, and the coefficients $v_i$ are the VEVs of the corresponding
fields.~\footnote
%--------------------------------------------------------------------------
{We must choose $v_l \ll V$ for the majoron not to contribute to
the invisible $Z^0$ width.}
%--------------------------------------------------------------------------

If we take the neutralino as LSP, the breaking of $R$ parity enforces the
process $\chi \rightarrow \nu_L +J$ as the main decay channel of $\chi\,$.
The decay lifetime is given by \cite{bemava}
%--------------------------------------------------------------------------
\begin{equation} \label{majlife}
\tau = 6 \cdot 10^{-\,26} \:\mbox{sec}\; \alpha_J^{-\,1} \:\left(
\frac{50 \,\mbox{GeV}}{m_\chi} \right) \;\;.
\end{equation}
%--------------------------------------------------------------------------
The coupling constant for this interaction ($\alpha_J = f^2 / 4 \pi$)
depends on the mixing angles between the components of the neutralino that
follow by diagonalizing the neutralino mass matrix \cite{bemava}. In the
present discussion we choose for definiteness $\chi = {\tilde Z}\,$. In
this case we have
%--------------------------------------------------------------------------
\begin{equation}
\alpha_J = \frac{\alpha_2}{2\:\mbox{cos}^2 \vartheta_W} \:
\frac{v_l^2}{v_r^2} \;\;.
\end{equation}
%--------------------------------------------------------------------------

We see from Eq.~(\ref{majlife}) that, in order to have have a late decay,
a very tiny violation of $R$ parity is needed. Despite such a small value
could seem rather unnatural, it is not inconceivable that it could arise
as a residual effect of gravitation \cite{bemava}. 

For $m_{\tilde Z} \simeq 50\:$GeV the main annihilation channel of the
zino is ${\tilde Z} \, {\tilde Z} \,\rightarrow\, f \,{\bar f}$ through
sfermion exchange. This gives \cite{ehnos}
%--------------------------------------------------------------------------
\begin{equation}
m\:Y_{\tilde Z} = 1.5 \cdot 10^{-\,7} \:\mbox{GeV}\: \left(
\frac{50\,\mbox{GeV}}{m_{\tilde Z}} \right)^2 \: \left( \frac{m_{\tilde
f}}{\mbox{TeV}} \right)^4 \;\;,
\end{equation}
%--------------------------------------------------------------------------
where $m_{\tilde Z}$ is the mass of the sfermions.

The shift of $\lambda_{\rm eq}$ thus requires
%--------------------------------------------------------------------------
\begin{equation}
\tau = \left( 1.3 \cdot 10^8 \div \underline{2.9 \cdot 10^9} \right)
\:\mbox{sec}\: \left( \frac{m_{\tilde Z}}{50 \, \mbox{GeV}} \right)^4 \:
\left( \frac{\mbox{TeV}}{m_{\tilde f}} \right)^8 \;\;.
\end{equation}
%--------------------------------------------------------------------------
Eqs.~(\ref{upplim}) and (\ref{lowlim}) can be ``translated'' into the
bounds
%--------------------------------------------------------------------------
\begin{equation} \label{bound1}
\left( 250 \div \underline{5700} \right) \:\mbox{sec}\: < \tau < \left(
8.6 \cdot 10^{12} \div \underline{5.0 \cdot 10^{12}} \right) \:\mbox{sec}
\end{equation}
%--------------------------------------------------------------------------
on the zino lifetime. A more stringent upper bound comes from the non
observation of the neutrinos, Eq.~(\ref{neuobs}):
%--------------------------------------------------------------------------
\begin{equation} \label{bound2}
\tau < 2.6 \cdot 10^{10} \:\mbox{sec} \: \left(
\frac{50\,\mbox{GeV}}{m_{\tilde Z}} \right)^2\;\;.
\end{equation}
%--------------------------------------------------------------------------
The requirement imposed by BBN are instead always satisfied for $m_{\tilde
Z}$ not much greater than $50\:$GeV, since (see Fig.~\ref{fig2})~\footnote
%--------------------------------------------------------------------------
{for $m_{\tilde Z} \leq 50\:$GeV and for $v_r \sim 100\:$GeV we do not
have to worry either about electromagnetic cascades produced by other
decay channels of ${\tilde Z}\,$. Indeed, the branching ratio for the
decays of the zino into photons or charged particles is \cite{bemava}
\begin{equation}
{\rm B.R.} \simeq 8 \cdot 10^{-\,5} \: \left( \frac{v_r}{100\,\mbox{GeV}} 
\right)^2 \: \left( \frac{m_{\tilde Z}}{50 \, \mbox{GeV}} \right)^2\;\;.
\end{equation}
}
%--------------------------------------------------------------------------
\begin{equation}
f_{\nu \rightarrow \gamma} \simeq 3 \cdot 10^{-\,3} \: \left( 
\frac{m_{\tilde Z}}{50 \, \mbox{GeV}} \right) \: 
\left( \frac{10^4\:\mbox{sec}}{\tau} \right) \;\;.
\end{equation}
%--------------------------------------------------------------------------
We thus have the final bounds
%--------------------------------------------------------------------------
\begin{equation}
\left( 250 \div \underline{5700} \right) \:\mbox{sec}\: < \tau < 2.6 
\cdot 10^{10} \:\mbox{sec} \: \left(
\frac{50\,\mbox{GeV}}{m_{\tilde Z}} \right)^2 \;\;,
\end{equation}
%--------------------------------------------------------------------------
which can be ``translated'' into the limit
%--------------------------------------------------------------------------
\begin{equation}
\left( 0.5 \div \underline{0.8} \right) \:\mbox{TeV}\; \left( 
\frac{m_{\tilde Z}}{50\, \mbox{GeV}} \right)^{3/4} < m_{\tilde f} < 5
\:\mbox{TeV}\; \left( \frac{m_{\tilde Z}}{50\, \mbox{GeV}} \right)^{1/2}
\end{equation}
%--------------------------------------------------------------------------
for the sfermion mass, and into
%--------------------------------------------------------------------------
\begin{equation}
5 \cdot 10^{-\,18} \: \left( \frac{m_{\tilde Z}}{50\, \mbox{GeV}}
\right)^{1/2} < f < \left( 5 \cdot 10^{-\,14} \div \underline{1 
\cdot 10^{-\,14}} \right) \: \left( \frac{50\, \mbox{GeV}}{m_{\tilde Z}} 
\right)^{1/2}
\end{equation}
%--------------------------------------------------------------------------
for the coupling $f = \sqrt{4\,\pi\,\alpha_J}\,$. From the smallness of
$f$ in this allowed range, production of Majoron in stars \cite{raf,dss},
as well in laboratory experiments, result in unobservable effects.

%--------------------------------------------------------------------------
\section{$m \:<\: {\rm MeV}\:$ Neutrinos}
%--------------------------------------------------------------------------

For light neutrinos the abundance is mass independent \cite{ket} 
%--------------------------------------------------------------------------
\begin{equation} 
Y_\nu = 3.88 \cdot 10^{-\,2}
\;\;\;.\end{equation}
%--------------------------------------------------------------------------
With this value, Eq.~(\ref{kimeq}) is replaced by~\footnote
%--------------------------------------------------------------------------
{As in the previous section, the underlined value is referred to the $h=1$
case, while the other one to $h=0.4\;$. In the case of a decaying neutrino
we need a slightly greater amount of density energy in the decay, since in
the thermal background there are only $2$ neutrinos left \cite{kim}.}
%--------------------------------------------------------------------------
\begin{equation} \label{neuk}
\left( \frac{m_\nu}{{\rm MeV}} \right)^2 \: \left(
\frac{\tau}{{\rm sec}} \right) = 2.52 \cdot 10^3 \div \underline{5.39
\cdot 10^4}
\;\;\;.\end{equation}
%--------------------------------------------------------------------------
Since neutrinos are light and since we consider decays into invisible
particles, we only have the general bounds set in Sec.\ II. We thus
consider neutrinos with masses
$$\left( 15 \div \underline{93} \right) \:{\rm eV}\: < \, m_\nu \,<\, 1
\:{\rm MeV}\;\;\;,$$
corresponding to lifetimes
%--------------------------------------------------------------------------
\begin{equation} \label{limtau}
\left( 2.5 \cdot 10^3 \div \underline{5.4 \cdot 10^4} \right) \:{\rm sec}\:
< \: \tau \: < \: \left( \underline{6.2 \cdot 10^{12}} \div
1.1 \cdot 10^{13} \right) \:{\rm sec}
\;\;\;.\end{equation}
%--------------------------------------------------------------------------

In the following we apply this scheme to three different scenarios
involving decaying neutrinos. For further details see also \cite{val},
where a more general discussion about decaying neutrinos (which however
does not tackle the particular problem of the shift of the equivalence
scale) is included. 

%--------------------------------------------------------------------------
\subsection{Neutrinos in left-right symmetric models}
%--------------------------------------------------------------------------

\begin{figure}[h!]
\begin{center}
\epsfig{bbllx=130pt,bblly=340pt,bburx=470pt,bbury=650pt,height=8truecm,
        figure=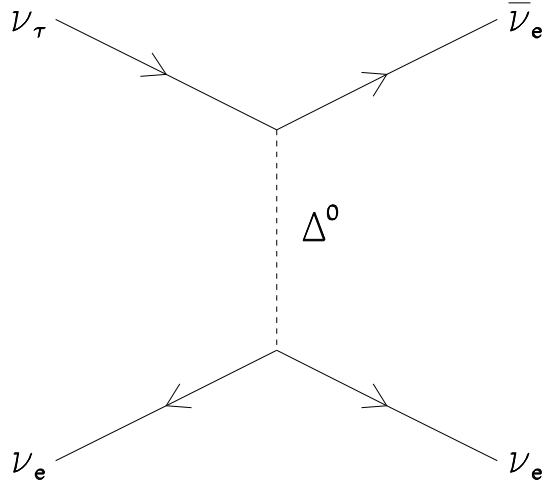}
\caption{\label{fig3}
\footnotesize Neutrino decay via isospin triplet.}
\end{center}
\end{figure}

In models with left-right symmetry there exist natural isospin triplets
which can mediate neutrino decay \cite{rs} (see Fig.~\ref{fig3}). 
%=========================================================================
%=========================================================================
The lifetime is given by
%--------------------------------------------------------------------------
\begin{equation} \label{nlrt}
\tau \left( \nu_\tau \rightarrow \nu_e \, \nu_e \, {\bar \nu}_e \right) =
\left( \frac{ m_{\Delta_L}^0}{m_W} \right)^4 \: \left( \frac{m_\mu}
{m_{\nu_\tau}} \right)^5 \: u_{\nu_\tau \, \nu_e}^{-\,2} \cdot \tau
\left( \mu \rightarrow \nu_\mu \, e \, {\bar \nu}_e \right) 
\;\;\;,\end{equation}
%--------------------------------------------------------------------------
where $u_{\nu_\tau \, \nu_e}$ is the mixing between the two families
and where the mass of the triplet is naturally of the order at which the
left-right symmetry is broken \cite{rs}:
%--------------------------------------------------------------------------
$$ m_{\Delta_L}^0 \simeq v_r > \:{\rm TeV}\;.$$ 
%--------------------------------------------------------------------------
Defining $\epsilon = m_W / v_r$ we thus have $\epsilon < 10^{-\,1}\:$.
Furthermore it must be $\epsilon > 10^{-\,4}$ or radiative decay would
prevail \cite{sv}.
%--------------------------------------------------------------------------
From Eq.~(\ref{neuk}) we get
%--------------------------------------------------------------------------
\begin{equation}
\left( \frac{m_\mu}{m_{\nu_\tau}} \right)^5 = \left( \underline{1.98 \cdot 
10^{-\,2}} \div 42.0 \right) \: \left( \frac{\tau}{{\rm sec}}
\right)^{5/2}
\;\;\;,\end{equation}
%--------------------------------------------------------------------------
and thus, from Eq.~(\ref{nlrt}),
%--------------------------------------------------------------------------
\begin{equation}
\frac{\tau}{{\rm sec}} = \left( 497 \div \underline{8.19 \cdot 10^4} 
\right) \: \left( u_{\nu_\tau\,\nu_e} \right)^{4/3} \:
\epsilon^{8/3}
\;\;\;.\end{equation}
%--------------------------------------------------------------------------
If we limit ourselves to $m \:<\:$ MeV neutrinos [and hence we take
into account the lower bound set in Eq.~(\ref{limtau})], even in case of
maximal mixing ($\:u_{\nu_\tau\,\nu_e}=1\,$) we have
%--------------------------------------------------------------------------
\begin{equation}
\tau > \left( 2.5 \cdot 10^3 \div \underline{5.4 \cdot 10^4} \right)
\:\mbox{sec} \,\rightarrow\, \epsilon > 1.8 \div \underline{0.9}\;\;,
\end{equation}
%--------------------------------------------------------------------------
which are above the allowed values.

%--------------------------------------------------------------------------
\subsection{Neutrinos in models with global symmetries}
%--------------------------------------------------------------------------

We now consider a theory with a global symmetry spontaneously broken. The
breaking induces the effective derivative coupling
%--------------------------------------------------------------------------
\begin{equation}
\Delta {\cal L} = V^{-\,1} \, j_\mu \, \partial^{\,\mu} \, \phi
\;\;\;,\end{equation}
%--------------------------------------------------------------------------
where $\phi$ is a (massless) goldstone scalar (or pseudoscalar) boson,
$j_\mu$ is the current associated to the fermions which couple to the
boson, and $V$ is the scale of symmetry breaking. 

Due to the very weakly interaction of these bosons with the ordinary
matter, decays $\nu \rightarrow \nu^\prime \, \phi$ have proven very
feasible for the shift of the equivalence scale \cite{wgs,kima}. Here we
comment on two of the most popular types of goldstone bosons in which a
heavy neutrino can decay into, i.e.\ familons and majorons. 

Familons arise from the spontaneous breaking of an horizontal symmetry
among the different generations of fermions \cite{wil,fam,gnp}. The
coupling familon-neutrinos can be written
%--------------------------------------------------------------------------
\begin{equation} \label{fameq}
\Delta {\cal L} = V_{\,\nu^\prime \,\nu}^{-\,1} \, {\bar \nu}^\prime \,
\gamma_\mu \,\nu \, \partial^{\,\mu} \, \phi
\;\;\;.\end{equation}
%--------------------------------------------------------------------------

If the natural assumption of universality between the couplings of the
familon to neutrinos and to charged leptons is made, the scale of symmetry
breaking must be taken $V > 8\cdot 10^{10} \:{\rm GeV}\:$ in order to
avoid problems with stellar energy loss due to familon bremsstrahlung of
the electrons in the stellar medium \cite{raf}.~\footnote
%--------------------------------------------------------------------------
{We are implicitly assuming that the lightest generation is involved in
the family symmetry.}
%--------------------------------------------------------------------------

Eq.~(\ref{fameq}) leads to the neutrino lifetime
%--------------------------------------------------------------------------
\begin{equation}
\tau \left( \nu \rightarrow \nu^\prime \, \phi \right) = 3 \cdot 10^6
\:{\rm sec}\:\left( \frac{V}{10^{10} \:{\rm GeV}} \right)^2 \: \left(
\frac{{\rm MeV}}{m_{\nu'}} \right)^3
\;\;\;.\end{equation}
%--------------------------------------------------------------------------
From Eq.~(\ref{neuk}) it follows
%--------------------------------------------------------------------------
\begin{equation}
\frac{V}{10^{10} \:{\rm GeV}} = \left( 0.21 \div \underline{2.0}
\right) \: \left( \frac{{\rm sec}}{\tau} \right)^{1/4}
\;\;\;.\end{equation}
%--------------------------------------------------------------------------
Limiting ourselves again to $m \:<\:$ MeV neutrinos, from 
Eq.~(\ref{limtau}) we have
%--------------------------------------------------------------------------
\begin{equation}
1.3 \cdot 10^{-\,4} \div \underline{1.1 \cdot 10^{-\,3}} < 
\frac{V}{10^{10} \:{\rm GeV}} < 0.03 \div \underline{0.13} \;\;.
\end{equation}
%--------------------------------------------------------------------------
The upper limits are below the allowed value for $V$ stated above.

\vspace{1cm}

The stellar energy loss bound can be considerably weakened if we consider
a goldstone boson which directly couples only to neutrinos. This is the
case for the majoron \cite{cmp} which arises in models with the
spontaneous breaking of the lepton symmetry ($L$). 

The Gelmini-Roncadelli model \cite{cmp} makes use of only left-handed
neutrinos and has lepton number spontaneously broken by the VEV of an
SU(2)$_{\rm L}$ Higgs triplet. Nowadays this option is ruled out by the
invisible width of $Z^0\,$.

We consider, instead, Majoron models {\em \`a la\/}
Chikashige-Mohapatra-Peccei \cite{cmp}, with right-handed neutrinos and
$L$ broken by the VEV of an isosinglet scalar $\phi\,$. This scalar
carries $L=\;-2$ and is coupled to two right-handed neutrinos through a
Yukawa interactions. When $\phi$ acquires a VEV, $\langle \phi \rangle =
V$, L breaks and a Majorana mass term is recovered. In addition to this
term, a Dirac mass term couples left-handed and right-handed neutrinos via
the usual Yukawa interaction with the standard model higgs doublet. If $V$
is taken much bigger than the electroweak scale $M_d$, the diagonalization
of the neutrino mass matrix splits the eigenvectors into heavy (mainly
$\nu_R\,$, with mass of the order of V) and light (mainly $\nu_L\,$, with
mass of the order $M_d^2/V$) neutrinos (see-saw mechanism \cite{sees}).

However, this minimal model gives for the decay $\nu_L\rightarrow
\nu_L^\prime \; \phi$ a lifetime bigger than the age of the universe,
since the leading order term vanishes for flavor changing processes
\cite{sv}. This problem can be circumvented in non minimal models (see
\cite{grr} and references therein). For natural values of the couplings,
these models give lifetimes of the order \cite{wgs}
%--------------------------------------------------------------------------
\begin{equation} \label{majo}
\tau \left( \nu \rightarrow \nu^\prime \, \phi \right) = 6 \cdot 10^4
\:{\rm sec}\: \left( \frac{V}{10^{10} \:{\rm GeV}} \right)^2 \: \left(
\frac{{\rm MeV}}{m_{\nu}} \right)^3
\;\;\;.\end{equation}
%--------------------------------------------------------------------------
From Eqs.~(\ref{neuk}) and (\ref{majo}) we get
%--------------------------------------------------------------------------
\begin{eqnarray} \label{majres}
m_\nu &=& \left( \underline{0.1} \div 2 \right) \,\mbox{keV}\: \left(
\frac{{\rm V}}{10^8 \:{\rm GeV}} \right)^2 \nonumber\\
\tau &=& \left( 4 \cdot 10^8 \div \underline{4 \cdot 10^{12}} \right)
\,\mbox{sec}\: \left( \frac{10^8 \:{\rm GeV}}{{\rm V}} \right)^4
\nonumber\\
M_d &=& \left( \underline{3} \div 14 \right) \:\mbox{GeV}\: \left(
\frac{{\rm V}}{10^8 \:{\rm GeV}} \right)^{3/2}
\;\;\;.\end{eqnarray}
%--------------------------------------------------------------------------

The main bounds on the scale $V$ come in this case from majoron emission
in supernovae explosions, and in particular from the neutrino pulse
observation from SN1987A. Although this bounds depend on the model used
and the discussion about them is not yet completely set \cite{gnp,raf},
from \cite{cs} we see that values $V >$ TeV can be reasonably allowed.
From Eq.~(\ref{majres}) we see that scales $V$ about $(10^8 \,\div\,
10^9)$ GeV (values well above the SN limits) are suitable to obtain the
shift of the equivalence scale.

%--------------------------------------------------------------------------
\section{Conclusions}
%--------------------------------------------------------------------------

In this paper we have considered several models with $\Omega_{\,{\rm CDM}
\,+\, {\rm baryons}}\, = \, 1 $ plus the presence of a late decaying
particle in order to match the observed spectrum of fluctuations. 

While having such late decaying particles is not unnatural in the most
popular extensions of the standard model, our analysis shows that in most
cases these models are severely constrained by cosmological and
astrophysical considerations. The main reason is that in the scheme we are
considering the decaying particle must dominate over the cold dark matter
and the relic radiative background before it decays. Thus the decay
products have high abundance and easily enter in conflict with bounds
imposed, for example, by their non observation or by nucleosynthesis.

The stiffness of the problem becomes clear when one considers decays with
the production of photons (or charged particles). In this case severe
bounds force the photon energy density to be much less than the one needed
to shift the equivalence scale. As a consequence, all the decays that
give rise to the shift must have a negligible production of photons.

Less constrained but still delicate is the case when high energy neutrinos
are produced. In addition to the bounds imposed by their non observation,
we cannot neglect the possibility that they scatter off background
thermal neutrinos, with consequent production of photons. We have seen
that these bounds are strictly related to the energy of each neutrino.
Thus --- for equal amount of energy density of the produced neutrinos ---
bigger abundances and lower energies of the single particles are
preferred.

Among the models studied, we focused in particular on supersymmetric
extensions of the standard model. We have split the discussion in models
with conserved and with broken $R$ parity.

In the case of conserved $R$ parity the main problems come from the
necessity of delaying the decay time. Rather than fine tuning the masses
(we saw that, in typical cases, masses must differ less than one part
over  $10^{13}\:$) we have studied models with very weakly coupled
particles, like gravitinos or axinos. However, we saw that even in this
case many difficulties arise.

If we instead choose to break $R$ parity, the natural candidate for the
decaying particle is the LSP. Models where $R$ is explicitly broken via
soft term can hardly accomplish the shift of the equivalence scale, since
avoiding with the latter the production of photons or charged particles
is very difficult. The situation improves in schemes where $R$ parity is
spontaneously broken by the $VEV$ of an isosinglet, where decays like
neutralino into neutrino and majoron can easily implement the shift of the
equivalence scale.

Although we have presented particular processes in some detail, we have
tried to maintain the discussion on the most possible general ground,
underlying the main difficulties that each different class of models faces
and (when possible) some ways to overcome them.

Given that evidences for $\:\Omega_{\rm matter}\:<\:1$ need more solid
confirmation, the option $\:\Omega_{\rm matter}\:=\:1$ with a delayed
equivalence scale remains still interesting. In case future observational
work would favor this latter options, we hope that this work will be a
basis for further more detailed investigations.

In particular, we think that important indications will come out from the
observation of the acoustic peaks of the CBR \cite{epb}. An analysis of the
peaks produced by the decays that we analyzed here could constitute an
important complement to the present work.

%--------------------------------------------------------------------------
\section{Acknowledgments}
%--------------------------------------------------------------------------

We are grateful to S.\ Borgani, E.\ Nardi, E.\ Pierpaoli, and A.\ Riotto
for useful and lively discussions and to V.S.\ Berezinsky for interesting
comments. In particular, we would like to thank S.\ Bonometto for pointing
out some incorrectness in the first version of this paper. The work of
D.M.\ is supported by an INFN Post-Doc fellowship. This work is partially
supported by the EEC TMR network ``Beyond the Standard Model'', contract
no.\ FMRX-CT96-0090.

%--------------------------------------------------------------------------

%--------------------------------------------------------------------------

\end{document}